\documentclass[namedreferences]{SolarPhysics}
\usepackage[optionalrh]{spr-sola-addons} 
\usepackage{graphicx}
\usepackage{color}           
\usepackage{url}             

\newcommand{\arcsec}{{$^{\prime\prime}$}}

\newcommand{\aap}{    {\it Astron. Astrophys.}}

\newcommand{\apj}{    {\it Astrophys. J.}}
\newcommand{\apjl}{   {\it Astrophys. J. Lett.}}

\newcommand{\mnras}{  {\it Mon. Not. Roy. Astron. Soc.}}

\newcommand{\solphys}{{\it Solar Phys.}}

\begin{document}

\begin{article}

\begin{opening}

\title{High-Energy Emission from a Solar Flare in Hard X-rays and Microwaves}

\author{M.R.~\surname{Kundu}$^{1}$\sep
        V.V.~\surname{Grechnev}$^{2}$\sep
        S.M.~\surname{White}$^{1}$\sep
        E.J.~\surname{Schmahl}$^{1,3}$\sep
        N.S.~\surname{Meshalkina}$^{2}$\sep
        L.K.~\surname{Kashapova}$^{2}$}

\runningauthor{Kundu et al.}
 \runningtitle{High-Energy Emission}

\institute{${}^{1}$Astronomy Department, University of Maryland, College
Park, MD 20742 email: \url{kundu@astro.umd.edu} email: \url{ed@astro.umd.edu}
email: \url{white@astro.umd.edu} \\
 ${}^{2}$Institute of Solar-Terrestrial Physics, Irkutsk 664033,
Russia email: \url{grechnev@iszf.irk.ru} \\
 ${}^{3}$Lab for Astronomy and Solar Physics, NASA Goddard Space
Flight Center, Greenbelt, MD 20771
 }

\begin{abstract}

We investigate accelerated electron energy spectra for different
sources in a large flare using simultaneous observations obtained
with two instruments, the Nobeyama Radio Heliograph (NoRH) at 17
and 34 GHz, and the Reuven Ramaty High Energy Solar Spectroscopic
Imager (RHESSI) at hard X-rays. This flare is one of the few in
which emission up to energies exceeding 200 keV can be imaged in
hard X-rays. Furthermore, we can investigate the spectra of
individual sources up to this energy. We discuss and compare the
HXR and microwave spectra and morphology. Although the event
overall appears to correspond to the standard scenario with
magnetic reconnection under an eruptive filament, several of its
features do not seem to be consistent with popular flare models.
In particular we find that (1) microwave emissions might be
optically thick at high frequencies despite a low peak frequency
in the total flux radio spectrum, presumably due to the
inhomogeneity of the emitting source; (2) magnetic fields in
high-frequency radio sources might be stronger than sometimes
assumed; (3) sources spread over a very large volume can show
matching evolution in their hard X-ray spectra that may provide a
challenge to acceleration models. Our results emphasize the
importance of studies of sunspot-associated flares and total flux
measurements of radio bursts in the millimeter range.

\end{abstract}


\end{opening}


\section{Introduction}

Energetic electrons accelerated to energies of tens and hundreds
of keV can be observed through microwave and hard X-ray (HXR)
emissions from the solar corona. Imaging observations are
important to study the origin of energetic electrons in large
flare events, which in turn can be used to test flare models and
other related theoretical issues. Two dedicated solar imaging
instruments most important for studies of solar flares are at
present available --- one in X-rays and gamma-rays by the Reuven
Ramaty High Energy Solar Spectroscopic Imager (RHESSI,
\opencite{Lin2002}) and the other in microwaves by the Nobeyama
Radioheliograph (NoRH, \opencite{Nakajima1994}) at 17 and 34 GHz.
NoRH is capable of imaging signatures of microwave emitting
electrons in flaring sources. At 17 GHz it measures both Stokes
$I$ and $V$, and at 34 GHz Stokes $I$ alone, with good sensitivity
and spatial resolution of $\approx \,$10\arcsec\ and $\approx
\,$5\arcsec, respectively at the two frequencies. Signatures of
hard X-ray emitting electrons are mapped by RHESSI. RHESSI's
primary objective is the study of energy release and particle
acceleration in solar flares. This is accomplished by imaging
spectroscopy of solar hard X-rays and gamma-rays over a 3 keV to
17 MeV energy range with energy resolution of $\approx\,$1 keV,
time resolution of $\approx\,$4~s or better and spatial resolution
as high as 2.3\arcsec.

Non-thermal microwave emission during large solar flares is
produced by the gyro\-syn\-chro\-tron mech\-anism which involves
coronal magnetic fields of at least a few hundred gauss and
electrons of hundreds of keV and higher energy. Hard X-ray
emission, on the other hand, is mainly produced by bremsstrahlung
from precipitating electrons of tens to hundreds of keV energies.
The two different methods of mapping energetic flare electrons
therefore complement each other, and provide good means of testing
flare-related concepts which have been abundant in the recent
literature. The major hard X-ray flux is emitted by precipitating
electrons striking a thick target, whereas microwaves are emitted
by electrons gyrating in magnetic fields, both precipitating and
trapped in coronal magnetic tubes.

Several issues related to accelerated electrons in solar flares
are debated in the literature. First, it is not clear if a single
acceleration mechanism operates in a flare or different mechanisms
contribute (\textit{e.g.}, \opencite{WSW63};
\opencite{BogachevSomov2001}). Note that the possible presence of
different ``accelerators'' does not necessarily show up in the
shape of the electron spectrum \cite{BogachevSomov2007}. One
cannot also rule out the possibility that in an event with
repetitive acceleration/injection episodes part of an electron
population accelerated in the previous episode undergoes an
additional acceleration from basically the same mechanism.

One more question is related to the fact that the electron spectra
inferred from microwave observations at frequencies believed to be
optically thin appear to be harder than those inferred from HXR
data as initially shown by \inlinecite{Kundu94} and repeatedly
confirmed afterwards. Following the interpretation of
\inlinecite{MelnikovMagun98}, other researchers (\textit{e.g.},
\opencite{Silva2000}; \opencite{Lee2000}; \opencite{Takasaki2007})
suggest that this fact can be explained by the collisional
hardening of the electron spectra in magnetic traps.

Another possibility was proposed by \inlinecite{White2003} who
considered high-frequency radio emissions at 35 and 80 GHz in the
well-studied flare of 23 July 2002. From the analysis of the
microwave/millimeter and HXR data they concluded that the trapping
could not explain the difference between the electron indices
inferred from these emissions in that event. They were ``forced to
assume that ... the 35\,--\,80 GHz spectrum does not represent
optically thin emission'' due to contamination at high frequencies
by a component with a very high turnover frequency.  This event
required a very large number of emitting electrons with a hard
spectrum, which they indeed found in that event, up to
$10^{10}$~cm$^{-3}$ above 20~keV with a power-law index $\delta
\approx 4.5-5$. The conclusion of \inlinecite{White2003} suggests
that a similar situation might occur in other events; with a
lesser number of power-law electrons, the optically thick regime
could reach high radio frequencies if magnetic fields are strong.

Here we discuss the RHESSI HXR and NoRH microwave imaging
observations of the flare of 17 June 2003. The flare in question
was of class M6.8, and it was observed in AR~10386 (S08\,E58), a
$\beta\gamma\delta$-region, two days after its east-limb passage.
This flare produced a multitude of strong, isolated bursts seen in
HXR and microwaves, combining a few similar events occurring at
nearly the same place under similar conditions and promising
important information on accelerated electrons. This flare was
previously discussed by \inlinecite{Ji07} in the context of
motions of flare loops. In our paper, we address high-energy
emissions observed during this flare both in the microwave and
hard X-ray domains. Emissions exceeding 300 keV have been imaged
in very few flares, while 300\,--\,800 keV emission was well
pronounced during one of the peaks of the flare under discussion.
The sources of such high-energy emissions have been mapped in only
few events so far; the 17 June 2003 flare is one such event.

This event is exceptional in being spatially extended and
affording an opportunity to measure the HXR spectra of several
distinct features in several different light-curve peaks up to
several hundred keV, and to compare with radio spectra: this
allows us to address aspects of the acceleration mechanism.

We use the following notations for the power-law spectral indices:
$\delta$ corresponds to the electron number, $\gamma$ corresponds
to hard X-ray photons, $\alpha$ corresponds to microwave flux
densities, and $\alpha_{\mathrm{T}}$ corresponds to microwave
brightness temperatures.

\section{Observations}
\label{S-observations}

\begin{figure} 
  \centerline{\includegraphics[width=0.75\textwidth]
   {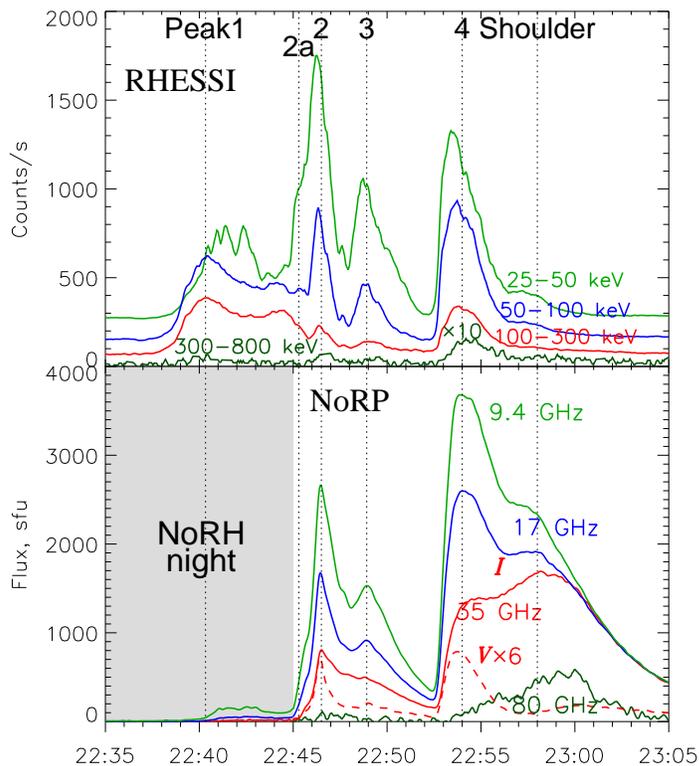}
  }
  \caption{RHESSI hard X-ray (top) and NoRP microwave (bottom) total
flux time profiles. The 300\,--\,800 keV band is magnified by a
factor of 10. Prominent peaks as well as a later shoulder are
denoted for convenience. RHESSI background levels are shifted to
show the bursts better. Radio light curves are shown for Stokes
$I$, and for Stokes $V$ at 35 GHz only (dashed; magnified by a
factor of 6).}
  \label{F-rhessi_norp}
  \end{figure}

The event was well observed by RHESSI and other instruments such
as the Transition Region and Coronal Explorer (TRACE,
\opencite{Handy1999}) at 1600 and 195~\AA. The Michelson Doppler
Imager (MDI, \opencite{Scherrer1995}) on SOHO has provided
magnetograms and continuum images close to the event occurrence.
RHESSI hard X-ray images are available from about 22:22 (all times
hereafter are UT). The event was also observed in soft X-rays by
GOES/SXI and in the H$\alpha$ line in Big Bear. The event started
with a filament eruption observed in extreme ultraviolet (EUV) and
H$\alpha$ \cite{Ji07}. The rise phase of the event occurred before
NoRH started observing for the day, so radio images are only
available from 22:45 onwards. In addition to NoRH, we also have
total flux data from the Nobeyama Radio Polarimeters (NoRP,
\opencite{Torii1979}; \opencite{Nakajima1985}) at seven
frequencies -- 1, 2, 3.75, 9.4, 17, 35, and 80 GHz. The HXR and
microwave light curves are shown in Figure~\ref{F-rhessi_norp}.
The impulsive rise in hard X-rays above 25 keV begins at 22:38,
with steepest rise at 22:39, and the first HXR maximum in the
50\,--\,300 keV bands occurs at about 22:40 (with several
sub-peaks in the 25\,--\,50 keV range).

\subsection{The Microwave and Hard X-ray Main Phase}
\label{S-main_phase}

In HXR time profiles in the main phase
(Figure~\ref{F-rhessi_norp}) we discern four distinct peaks 1, 2,
3, and 4. Peak~1 (22:39\,--\,22:43) is not observed by NoRH due to
night time. The spatial structures observed during an enhancement
labeled 2a (22:44:50\,--\,22:45:40) that passes into peak~2 differ
from those observed during peak~2 itself, and therefore we
consider it separately. Peak~2 (22:45:40\,--\,22:47:30), peak~3
(22:48\,--\,22:52), and peak~4 (22:53\,--\,22:57) are followed by
a plateau in the HXR emission profile which we call the shoulder
(22:57\,--\,23:00) discernible at 25\,--\,50, 50\,--\,100, and
100\,--\,300 keV. The hardest emission ($>$ 300 keV) is mostly
faint, becoming well pronounced only during peak~4.

The NoRP patrol time profiles confirm that the true radio onset in
microwaves was at 22:39. The radioheliograph at Nobeyama commenced
observing at 22:45, starting with the rise of HXR peak~2a. All
other peaks are observed by NoRP up to 80 GHz, including the
shoulder at the end of peak~4. The microwave time profiles
recorded in Nobeyama basically resemble the HXR records, but they
are smoother, and their maxima lag behind the HXR peaks by several
tens of seconds. This results in a larger overlap of peaks 2 and 3
with respect to hard X-rays. Note that the sharp drop in microwave
flux prior to peak 4 indicates that few non-thermal electrons were
trapped in the corona after peaks 2 and 3. Another oddity of the
light curves is the fact that the 100\,--\,300 keV HXR are
actually brightest during peak 1, but, \textit{e.g.}, the 17 GHz
radio emission is 50 times brighter in peak 4 than in peak 1. The
shoulder is more pronounced in microwaves than in hard X-rays,
even exceeding peak 4 at 35 and 80~GHz in intensity. The microwave
burst is very strong, reaching 3800 sfu at 9.4 GHz, 2600 sfu at 17
GHz, and 1700 sfu at 35 GHz. Also remarkable is the decay after
23:00, when the flux densities at 9.4, 17, and 35~GHz become
almost the same.

A strong emission up to $\approx\,$600 sfu is also recorded at 80
GHz. Measurements of the flux density at 80 GHz from NoRP records
are complicated by the following circumstances. A polarization
switch of the 80~GHz radiometer degraded for several years until
the problem was fixed on 23 June 2005. Accordingly, the flux
values at 80 GHz measured between June 1999 and 23 June 2005
gradually decreased with respect to their true values. To repair
the 80 GHz flux density, a time-dependent correction factor was
inferred from several calibrations (H.~Nakajima, 2006, private
communication):
\begin{eqnarray}
k_{\mathrm{cor}}\mathrm{(80\,GHz)}=[T_{\mathrm{\{year\}}}/1995.83]^{630}.
 \label{E-80ghz_corfactor}
\end{eqnarray}
The accuracy of the corrected total fluxes at 80~GHz within this
time interval is considered to be $\pm 40\%$, and polarization
measurements are not reliable. Data at 35 GHz and, especially,
80 GHz are affected by atmospheric absorption. Uncertainties
of the background level contribute to measurement errors. All
these factors decrease the measurement accuracy at high radio
frequencies.

The flux density at 9.4 GHz surpasses the fluxes at higher
frequencies during all peaks in the impulsive phase. The excess of
the 9.4 GHz emission is often considered as an indication that the
peak frequency of the flaring microwave sources is $< 17$~GHz, and
hence both 35 and 17~GHz emissions are believed to correspond to
the optically thin regime. However, this indication could be
misleading, as we show later.

\subsection{The Flare Configuration}

\subsubsection{Flare Ribbons and HXR Sources}

This flare occurred in a region which in white light (WL)
consisted of a complex of sunspots with strong umbrae and
penumbrae. We use TRACE 1600~\AA\ images to outline the flare
ribbons. The TRACE absolute pointing coordinates have an
uncertainty significantly larger than its spatial resolution of
$1^{\prime \prime}$\footnote{see
\url{http://trace.lmsal.com/Project/Instrument/cal/pointing.html}
and\\ \url{http://trace.lmsal.com/tag/}}, whereas those of RHESSI
are more accurate. For this reason, coordinates in all figures are
referred to the RHESSI coordinates. To co-align the TRACE and
RHESSI images, we compared the TRACE WL images with full-disk MDI
images. Solar rotation was compensated through a re-projection of
the continuum images to the same time (analogous routines were
performed with MDI magnetograms). Residual inaccuracies of a few
arc seconds are possible.

\begin{figure} 
  \centerline{\includegraphics[width=\textwidth]{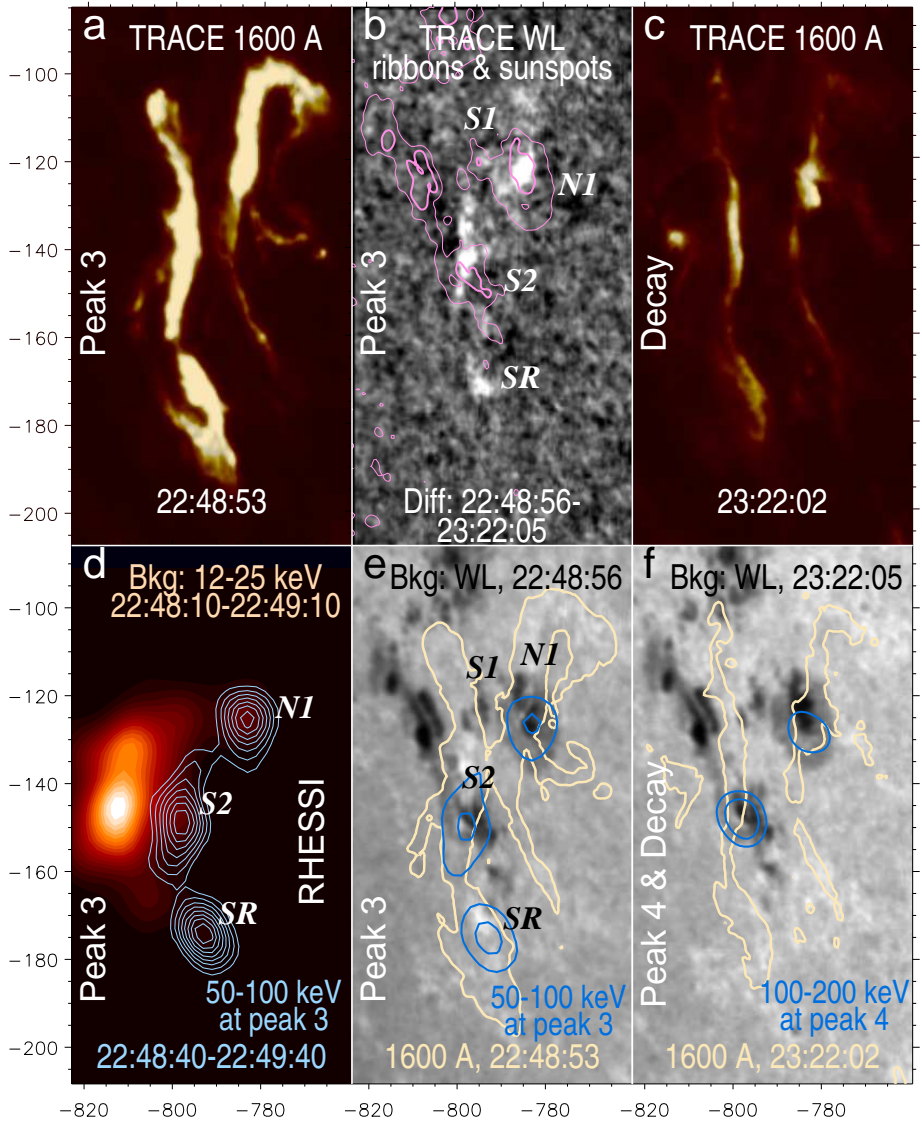} }
\caption{Flare ribbons, HXR sources, and sunspots. Ribbons were
observed by TRACE during peak 3 (a, b, e) and late in the decay
phase (c, f) in the 1600~\AA\ images (a, c) and WL ones (b, e, f).
Panel (b) shows a WL difference image with pink contours of the
sunspot umbrae and penumbrae. The lower row also shows RHESSI
images at peak 3 (d, e; 50\,--\,100 keV) and peak 4 (f;
100\,--\,200 keV). Levels of blue contours in panels (e, f) are
40\% and 80\% of the maximum. Yellow contours in panels (e) and
(f) correspond to the closest 1600~\AA\ images (a and e). ``N1'',
``S1'', and ``S2'' denote major sunspots related to the flare site
according to their polarities, and ``SR'' denotes the southern
region of the flare. Axes show hereafter arc seconds from the
solar disk center according to the pointing of RHESSI and MDI.}
  \label{F-1600-wl}
 \end{figure}

Figure~\ref{F-1600-wl} outlines the flare configuration. The TRACE
1600~\AA\ images show the flare ribbons at peak 3 (a) and late in
the decay phase (c). Their contours are also overlayed on top of
TRACE WL images observed at nearly the same times (e, f). The
RHESSI 12\,--\,25 keV and 50\,--\,100 keV images observed at
peak~3 are presented in panel (d), and the contours of the same
50\,--\,100 keV images are shown on top of the TRACE WL image (e).
Panel (f) shows the TRACE WL decay-phase image as grayscale
background along with flare ribbons and oval black/white dashed
contours of the RHESSI 100\,--\,200 keV images observed at peak~4.

Although the configuration looks like a two-ribbon flare
\cite{Ji07}, the situation is not quite the classic scenario.
Unlike an ordinary flare, the ribbons in this event cross and
almost cover sunspots N1 and S2. The emission of the ribbons is
faintly visible in the WL image in Figure~\ref{F-1600-wl}e. Panel
(b) presents a WL difference image, in which these brightenings
are clearly visible. They are most likely due to the leakage of UV
emissions in the TRACE wide-band continuum channel; alternatively,
they may show weak white-light emission (\textit{cf.}
\opencite{Metcalf2003}; \opencite{Hudson2006}). The HXR sources
also appear to be located within sunspots (N1 and S2), overlapping
with their umbrae. Besides the main sources associated with
sunspots N1 and S2, there is an additional HXR source south of S2
denoted ``SR''.

\subsubsection{Overall Description of the Event}
 \label{S-flare_story}

\begin{figure} 
  \centerline{\includegraphics[width=\textwidth]{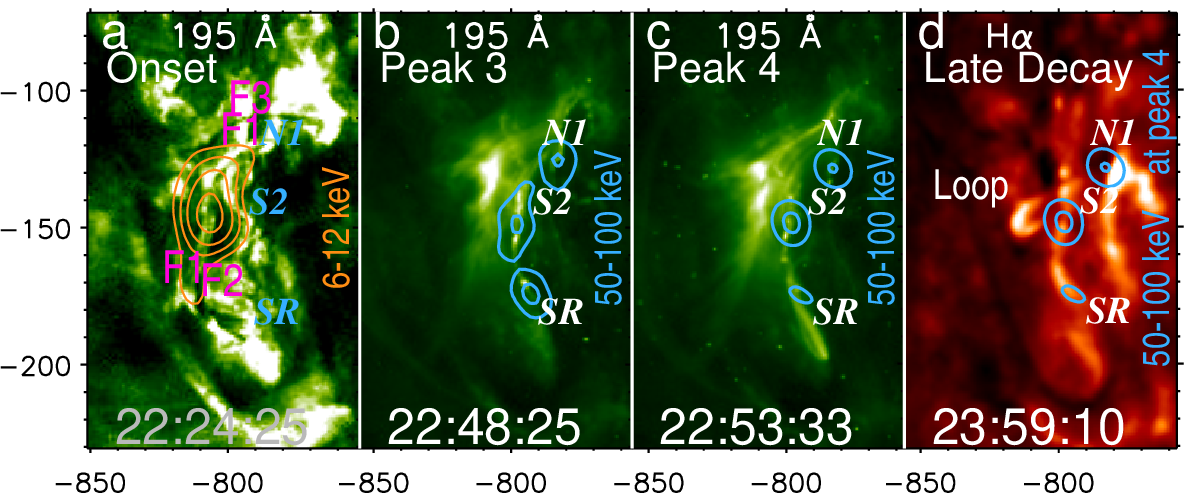} }
\caption{Coronal images of the event from the onset up to its late
decay: (a): activation of filaments, (b, c): peaks 3 and 4, (d):
late decay. Green background shows TRACE 195~\AA\ images in panels
(a\,--\,c), and red background shows a BBSO H$\alpha$ image in
panel (d). Contours show RHESSI images. Labels ``N1'', ``S2'',
``SR'' denote flare regions. Filaments visible in TRACE images are
labeled ``F1\,--\,F3'' in panel (a). A late-stage H$\alpha$ image
(d) shows a post-flare loop between N1 and S2.}
  \label{F-euv_arcade}
 \end{figure}

Having discussed the flare configuration and its main
characteristics, we now consider the development of the event from
its start up to the late decay. Coronal phenomena are shown by
TRACE 195~\AA\ images in Figure~\ref{F-euv_arcade}a\,--\,c as well
as an H$\alpha$ image in panel (d). A more detailed information is
presented by a movie \url{TRACE_RHESSI.mpeg} accompanying the
electronic version of our paper, which shows a movie composed of
the TRACE 195~\AA\ images overlayed with RHESSI 12\,--\,25 keV
(red) and 50\,--\,100 keV (green) contours.

A system of filaments (F1, F2, F3) covered the whole pre-event
region with their northern ends being rooted approximately between
N1 and S1 and the southern ends somewhere near SR. One of the
filaments (F1) activates at about 22:23, which is manifest in its
brightening, and starts to gradually rise
(Figure~\ref{F-euv_arcade}a). The activation probably also
involves filament F2. This time corresponds to the earliest
detectable increase of the soft X-ray flux recorded with GOES
monitors (the onset of the event in soft X-rays was reported to be
at 22:27). The brightening of the filament indicates heating up to
coronal temperatures, while RHESSI shows the presence of still
hotter plasmas in this region. The coronal X-ray source detectable
up to 25 keV is arranged along the brightening filament.

Then the filaments straighten, rise, and finally erupt (between
22:39:25 and 22:40:40), while their southern ends remain fixed in
the position close to the future southern flare region SR
(Figure~\ref{F-euv_arcade}b). Note that the first HXR peak occurs
simultaneously with the filament eruption. The flare sources N1
and SR are obviously associated with the positions of the ends of
pre-eruptive filaments.

The TRACE 195~\AA\ images obtained at peaks 3 and 4
(Figure~\ref{F-euv_arcade}b,~c) show a typical arcade of flare
loops arranged along the former position of the erupted filaments.
The arcade does not exhibit any conspicuous features. The most
noticeable are bright kernels in its base which coincide or almost
coincide with the HXR sources. H$\alpha$ images obtained late in
the decay phase of the event sequentially show cooling post-flare
loops. Remarkable is a loop between N1 and S2 caught in an image
shown in Figure~\ref{F-euv_arcade}d (the outer edge of its
northern leg is already getting dark, while the whole loop remains
semi-transparent).

\begin{figure} 
\centerline{\includegraphics[width=\textwidth]{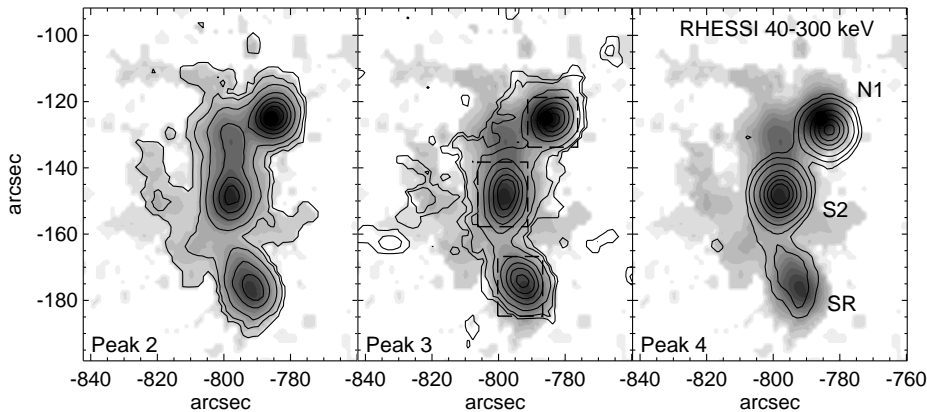}}
\caption{Hard X-ray images at each of the three main peaks in the
HXR light curve (contours). The images result from summing over
the 40\,--\,300 keV range. The background image in each panel is
the image for peak 2 so that changes in morphology from one peak
to the next can be seen. Contours are at 4, 8, 16, 32, 48, 64, and
80\% of the peak in each image. The resolution of the images is
9\arcsec. The middle panel shows the regions used for the HXR
spectra of sources S2, N1, and SR as dotted boxes.}
\label{F-hxr_images}
\end{figure}

\subsubsection{Hard X-Ray Morphology}

In hard X-rays, the time profile of the flare shows the usual
gradual behavior below 25 keV and multiple impulsive spikes at
higher energies. The images in the lower energy bands for
6\,--\,12 and 12\,--\,25 keV, most of which we do not present,
show that the flaring region consists of, with a few exceptions
(22:52:40\,--\,22:55:00), compact sources with apparent flux
maxima at $\sim$ 22:46\,--\,22:51 and 22:54\,--\,22:55. The
gradual profile continues until the end of RHESSI sunlight at
23:06. At these low energies, except from 22:52 to 22:55, one does
not see footpoint-like compact sources --- only a loop-like
structure. In the higher energy channels, above 40 keV, right from
the beginning the main flaring region is resolved into three or
more individual compact sources.

The HXR count rate profiles at 25\,--\,50 and 50\,--\,100 keV of
the impulsive phase show four peaks labeled 1, 2, 3, and 4 in
Figure~\ref{F-rhessi_norp}. Each of these has a different
morphology, spectrum, and temporal behavior. To obtain a coherent
perspective, we studied the morphology and spectra for each of
these. In each case we overlaid the HXR sources on an MDI
magnetogram made at the start of the flare. The main result that
comes out of this morphological study is that at lower energies
one sees the entire flaring loops, including in some cases the
loop tops. At higher energies one sees the footpoint sources. This
is especially true for peaks 2, 3, and 4 (in peak 1 we see HXR
emission close to sunspot S1 that does not recur in the later
peaks). An example is shown in Figure~\ref{F-1600-wl}d for peak 3.
Figure~\ref{F-hxr_images} also shows the morphology of peaks 2, 3
and 4 in the 40\,--\,300 keV range. Hard X-rays of hundreds of keV
are seen over most of the length of the flare ribbons, with an
extent of almost 60000 km north--south. Three main sources at S2,
N1 and SR are prominent in all three peaks, but there are changes
in morphology: in particular, peak 2 shows extended emission while
peak 4 is dominated by compact sources at S2 and N1. The HXR peak
over N1 shifts about 5\arcsec\ to the south-west as the flare
evolves from peak 2 to 3 to 4, whereas the S2 source appears at
the same location in each peak.

\subsubsection{Radio Sources}
\label{S-Radio_Sources}

NoRH images at 17 and 34 GHz were synthesized using the NRAO AIPS
package. The images were restored with gaussian beams of
full--width--half--maximum $12^{\prime \prime}$ at 17 GHz and
$8^{\prime \prime}$ at 34 GHz. At high flux levels the NoRH
calibration is established by using the NoRP total flux records.

\begin{figure} 
\centerline{\includegraphics[height=0.85\textheight]{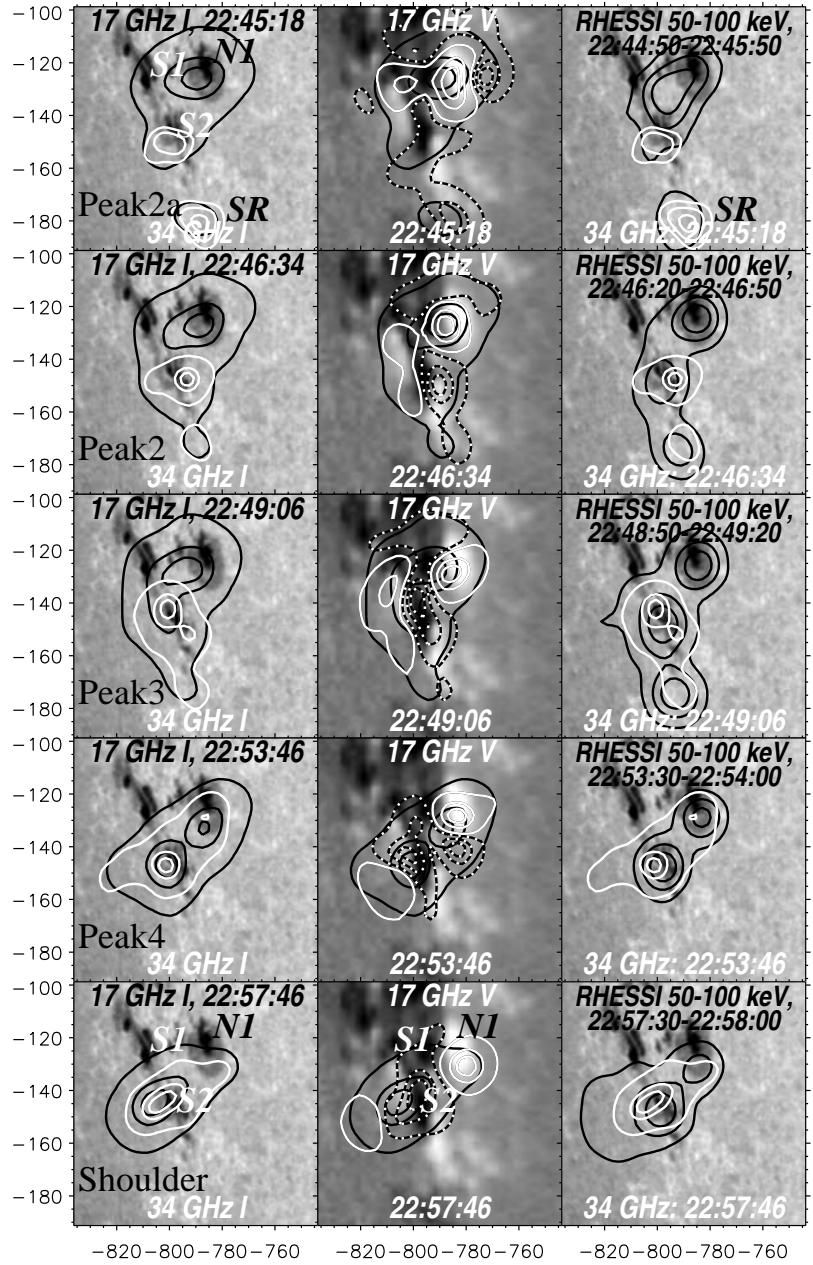}}
\caption{Flare morphology observed during peaks 2a\,--\,4 and
shoulder (consecutive rows). {Left column}: overlays of 17 GHz
(black) and 34 GHz (white) contour maps on a TRACE WL image
observed at 23:22:02. {Middle column}: 17 GHz Stokes $V$ maps
(white; solid positive, broken negative) and Stokes $I$ maps
(black) on an MDI magnetogram observed at 22:23 (bright N, dark
S). {Right column}: RHESSI 50\,--\,100 keV (black) and 34 GHz
contour maps on the same TRACE WL image as in the left column. }
 \label{F-overlay_set}
\end{figure}

The co-alignment of microwave sources with other images is not a
simple task. The coordinates of the NoRH images are generally
found by fitting for the position of the quiet solar disk, but
this is not always feasible when the flare flux much exceeds the
disk flux. Therefore, when the uncertainties of computations of
the solar disk's position are large, the positional accuracy of
the NoRH images is poor. They can also suffer from relative
shifts. The accuracy of the co-alignment can be improved by
referring to some features in other images, \textit{e.g.}, by
comparing the radio polarization with a magnetogram. Microwave
flare emissions are usually dominated by gyrosynchrotron emission
from power-law electrons, which can be significantly polarized in
the sense of the $x$-mode emission at optically thin frequencies.
Thus, the sign of the radio polarization coincides in this case
with the polarity of the magnetic field. At frequencies below the
turnover of the spectrum, where the emission is optically thick,
the radio polarization significantly decreases, and its sign
changes to the $o$-mode. Taking account of these circumstances, we
co-aligned the NoRH 17 GHz maps with the magnetogram and HXR
images. The residual inaccuracy might exceed a few arc seconds.
The NoRH images at 17 GHz are shown in Figure~\ref{F-overlay_set}
in the left column (Stokes $I$) on top of the TRACE WL image and
in the middle column (Stokes $I$ and $V$) on top of the pre-event
MDI magnetogram.

The accurate co-alignment of the 34 GHz images is still more
problematic. NoRH does not provide images of the polarized
component at 34~GHz. To find the correct positions of the 34 GHz
sources, one has to compare their shapes with those at 17 GHz or
HXR. The 34 GHz images are shown in Figure~\ref{F-overlay_set} in
the left and right columns. The absence of 34~GHz emission from
the region of the strongest 17~GHz emission over N1 in the three
upper rows is surprising, but after a careful analysis of the time
sequence of images at both 17 and 34 GHz and in HXR we are
confident that the co-alignment is correct to within a few arc
seconds.

\begin{figure} 
\centerline{\includegraphics[width=\textwidth]{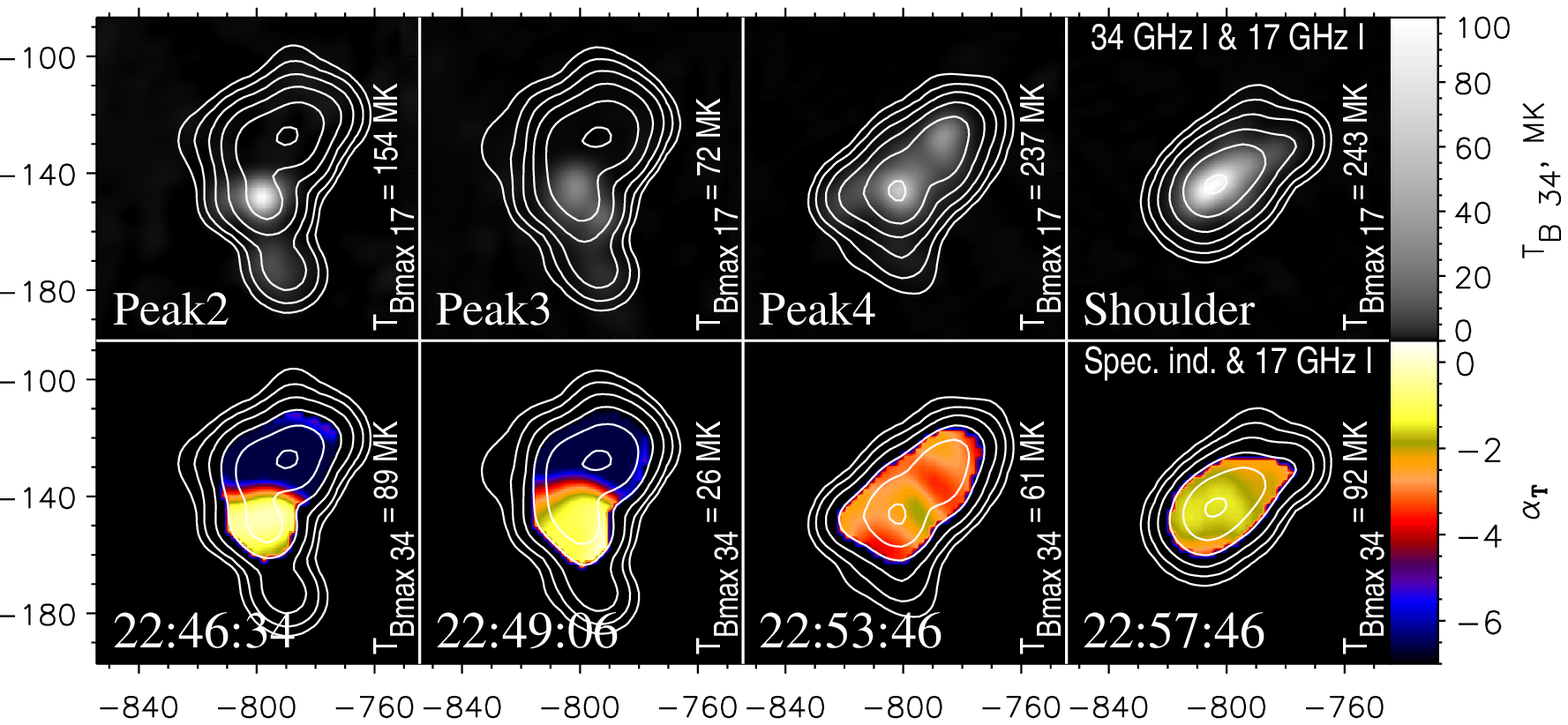}}
\caption{Microwave images and spectral indices. 34 GHz images (top
row) and spectral indices (bottom row, derived after convolving the
34 GHz images to the 17 GHz resolution) both overlaid by contours
of 17 GHz images. Contour levels in each image are at 0.9 of its
maximum divided by powers of 3. Scale bars on the right quantify
the grayscale and color representations. The maximum brightness
temperatures over each image are specified in the upper row for 17
GHz and in the lower row for 34 GHz.}
 \label{F-microwave_indices}
\end{figure}

Figure~\ref{F-overlay_set} shows microwave sources at 17 and 34
GHz (NoRH) and RHESSI 50\,--\,100 keV images on top of the
decay-phase TRACE image (left and right columns) and an MDI
magnetogram (middle column). The maximum brightness temperatures
in the images are shown in Figure~\ref{F-microwave_indices} (for
17 GHz in the top row and for 34 GHz in the bottom row). The areas
of the microwave sources measured at peak~4 are as follows. The
areas of the main 17 GHz sources are 207 arcsec$^2$ ($1.1 \times
10^{18}$~cm$^{2}$) in sunspot N1 and 162 arcsec$^2$ ($8.8 \times
10^{17}$~cm$^{2}$) in sunspot S2; the areas of the 34 GHz sources
are 36 arcsec$^2$ ($2 \times 10^{17}$~cm$^{2}$) in sunspot N1 and
108 arcsec$^2$ ($5.9 \times 10^{17}$~cm$^{2}$) in sunspot S2.

Starting from the onset of observations in Nobeyama, the 17 GHz
emission is dominated by sunspot N1; some contribution from S1 and
S2 is also detectable. A similar picture is shown by hard X-rays.
In addition, there is a detectable HXR emission from the southern
region SR. Unlike the 17 GHz emission, the 34 GHz sources are
concentrated in sunspot S2 and SR. From the enhancement 2a to peak
2 and then to peak 3, the source above sunspot S2 increases in
intensity, while the HXR emission in S1 relatively decreases. At
peak 4 and later on, sunspot S2 dominates, and the strongest
emissions come from its umbra, while the sources in sunspot N1
shift south, into its penumbra.

During peak 2 the southernmost flare source denoted ``SR'' is well
pronounced in hard X-rays, but weak at 17 GHz, obviously due to
significantly weaker magnetic fields in this region (the microwave
intensity is determined by the magnetic field strength). Unlike
the sources above sunspots N1 and S2, the source in SR is
localized above a bipolar magnetic region. The magnetic field
strength in this region varies from $-770$~G to $+900$~G under the
HXR source.

The magnetic field strengths on the photosphere were measured from
full-disk MDI magnetograms. Note that the MDI magnetograms were
recalibrated late in 2007, which resulted in an increase of the
magnetic field strengths by a factor of about 1.7 (see
\url{http://soi.stanford.edu}). In addition, the position of the
active region far from the solar disk center (S08\,E58) causes a
projectional reduction of the magnetic field strength. Its
correction is generally questionable, because the direction of the
magnetic field vector might be different. However, we are dealing
with main flare sources associated with sunspots, where the
magnetic field is nearly radial, and a radialization correction
appears to be justified. We have done this by using the
\textit{zradialize} SolarSoftware routine. The radialization
factor is from 1.82 in N1 up to 1.93 in S2. The maximum magnetic
field strength measured from the projection-corrected magnetograms
is $+3080$~G and $-2120$~G in sunspots N1 and S2 associated with
major microwave sources, respectively. The maximum strength in
sunspot S1 is $-1750$~G.

The microwave polarization is of special interest. It corresponds
fairly well with the magnetogram. The degree of polarization in N1
initially reaches 50\% at peak 2a, and then mostly persists at a
level of about 30\%. The degree of polarization in S2 is
20\,--\,30\% throughout the event. However, the contours of the
polarization do not perfectly correspond to the magnetic polarity
everywhere. The most conspicuous are discrepancies at peak
2a\,--\,2 in the region of the main source (N1) and both north and
south of it. The polarization structure varies greatly at this
time. Similarly, discrepancies are observed at peak 4 and the
shoulder northeast and southwest of S2. It is not possible to
explain the discrepancies between the microwave polarization and
the magnetogram by insufficient spatial resolution only; certainly
there are changes from the $x$-mode to the $o$-mode emission. This
fact hints at a possibility that the 17 GHz emission might not be
optically thin.

To check for this possibility, we show in
Figure~\ref{F-microwave_indices} the 17 and 34 GHz Stokes $I$
images (top row) and a microwave spectral index computed from the
spatial distributions of the brightness temperatures at these
frequencies (bottom row). The hardest optically thin microwave
index $\alpha_{\mathrm T}$ is $-3.5$ for the hardest realistic
power-law index of the electron number spectrum $\delta = 3$
($\alpha_{\mathrm T} = 1.22-0.9\delta-2$, \opencite{DulkMarsh82}).
As we will see below from HXR spectra, in this event $\delta > 4$,
and $\alpha_{\mathrm T} < -4.4$. Thus, orange, brown, and yellow
regions are certainly optically thick at 17 GHz; moreover, it is
possible that blue regions only are optically thin.

A movie \url{overlay_wl_17.gif} accompanying the electronic
version of our paper shows the radio blob superimposed on the
sunspot complex. The main point here is that the flaring source at
17~GHz covers the sunspot umbrae considerably. This is one of the
important characteristics of this flare, and may give rise to some
of the peculiar features of the event.

From this morphological study one sees that strong X-ray and
microwave flare emissions are radiated by a few loops, some of
which are rooted in sunspots. A loop between N1 and S1 is
detectable during peaks 2 and 3 and, especially, enhancement 2a
before them. The southern region SR probably also has a loop
structure and is connected with N1. Emissions during peak 4 and
the shoulder are dominated by a loop between N1 and S2. Both the
microwave and HXR sources have similar structures at this time.
The most important point to note here is that the flaring source
during peak 4 occurs just above sunspot umbrae. While footpoints
mainly radiate at 34 GHz and in HXR, emissions of the whole loops
are also detectable at 17 GHz.

\begin{figure} 
\centerline{\includegraphics[width=\textwidth]{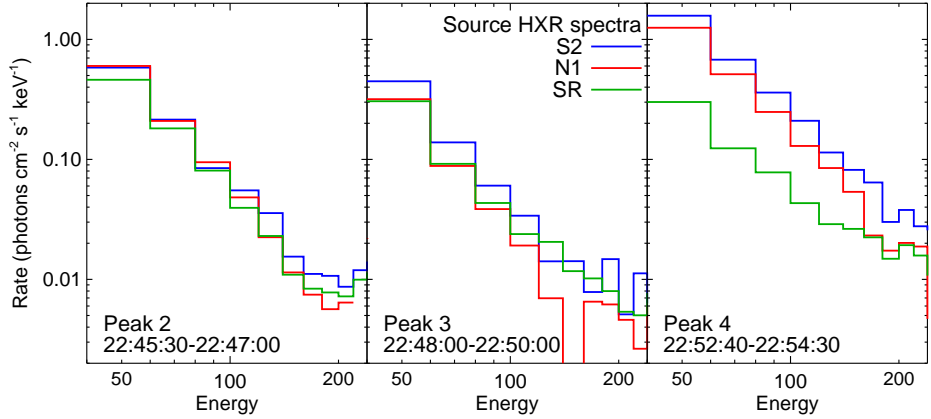}}
\caption{Hard X-ray spectra at each of the three main peaks in the
light curve for each of the three sources in the HXR images. These
are derived from image cubes made in 20 keV channels from 40 to
300 keV.
 } \label{F-hxr_spectra}
\end{figure}

\subsection{Spectral Data}
 \label{S-hxr_spectra}

In order to investigate the non-thermal electron energy spectra at
different locations in the flare volume, images were made at a
range of energies in a number of time intervals. Computations of
HXR spectra are complicated by several changes of the RHESSI
attenuator and decimation states, and possible time variation in
the background. We use both fits to the image data and the
standard background-subtracted data analysis to investigate the
HXR photon spectra. Images were deconvolved with both CLEAN and
Pixon methods for comparison. Since the non-thermal part of the
spectra is our major interest, we mostly consider spectra above
40~keV.

Spectra derived from the image cubes in the range 40\,--\,240 keV
with 20 keV bins are shown in Figure~\ref{F-hxr_spectra} for each
of the sources S2, N1, and SR at each of the peaks 2, 3, and 4.
Photon power-law indices $\gamma$ derived from these spectra, and
from the RHESSI front detectors (spatially integrated and
background-subtracted), are listed along with uncertainties in the
fits to the spectra in Table~\ref{tab}.

\begin{table}
\begin{tabular}{l|cccc}
\hline
Peak & S2 (middle) & N1 (north) & SR (south) & Spatially integrated \\
\hline
2: 22:45:30-22:47:00 & 3.1 $\pm$ 0.3 & 3.3 $\pm$ 0.3 & 3.0 $\pm$ 0.4 & 3.3 (280) 2.5 $\pm$ 0.2 \\
3: 22:48:00-22:50:00 & 3.3 $\pm$ 0.1 & 3.5 $\pm$ 0.1 & 2.8 $\pm$ 0.5 & 3.4 (210) 2.0 $\pm$ 0.2 \\
4: 22:52:40-22:54:30 & 2.7 $\pm$ 0.1 & 2.9 $\pm$ 0.2 & 2.1 $\pm$ 0.6 & 2.5 (120) 3.4 $\pm$ 0.2 \\
\hline
\end{tabular}
\caption{Power-law fits to the photon spectral index $\gamma$ of
individual sources in each of the three main peaks in the 17 June
2003 light curve derived from images in different energy bins,
together with the fit to background-subtracted 50\,--\,400 keV
spectra from the RHESSI front detectors. For the spatially
integrated spectra, the numbers are the results of a broken
power-law fit: the spectral index at energies below the break, the
break energy (keV) in parentheses, and the spectral index above
the break. Uncertainties in the fits to the break energies are
typically large (tens of keV). }\label{tab}
\end{table}

The fits to the spatially-resolved spectra assume a single power
law over the 40\,--\,240 keV range, while the fits to the
integrated spectra assume a broken power law over the range
50\,--\,400 keV. For peaks 2 and 3 the spectral break in the power
law is above 200 keV and the fitted spectral index below 200 keV
generally matches the fits to the spatially resolved spectra,
which are dominated by photons below 120 keV, while the fit above
the break gives a flatter spectrum. However for peak 4 the
spectral break is fitted to be at 120 keV: below this energy the
spectral index $\gamma$ is 2.5 while at higher energies it is 3.4.
The spatially resolved spectra at peak 4 are consistent with an
index of 2.5, as expected since they are dominated by the
40\,--\,120 keV photons. There is a suggestion that the spectrum
of source SR is somewhat harder than sources S2 and N1 during peak
4, but it is much weaker than those sources and the uncertainty in
the spectral index of SR is large (Table~\ref{tab}).

The general conclusion of Figure \ref{F-hxr_spectra} and
Table~\ref{tab} is that a given peak shows the same energy
spectrum in all three spatial locations, but it may differ from
one peak to the next: peaks 2 and 3 clearly have steeper spectra
than peak 4 (below 120 keV) in all three sources and in the
integrated spectra. This suggests either that the electron
acceleration mechanism has the same physical characteristics over
a large spatial scale (5 $\times$ 10$^5$ km) or a more localized
accelerator distributes electrons over the full volume. The
challenge for the first interpretation is the fact that all
sources show a flattening of their spectra in peak 4 after being
steeper in peaks 2 and 3: how can sources so far apart have their
characteristics change in the same way? For example, if
acceleration is due to stochastic acceleration by wave turbulence,
how is turbulence generated with identical properties over such a
large volume? On the other hand, a localized accelerator that can
distribute non-thermal electrons over a distance of 5 $\times$
10$^5$ km is difficult to reconcile with the usual picture of
post-flare loops in two-ribbon flares that are typically much
shorter than the ribbons and straddle the neutral line rather than
parallel the ribbons.

Peak 2 is of special interest, because at this peak we have
encountered a puzzling situation with no 34 GHz emission from the
region of the 17 GHz source at N1. Such a situation is possible if
the 17 GHz emission is thermal gyroresonance emission at low
harmonics of the gyrofrequency. However, the first panel of
Figure~\ref{F-hxr_spectra} clearly shows that the HXR spectrum of
N1 in the first peak does have non-thermal photons up to 200 keV
that must result from electrons with energies of order 500 keV or
more.

\subsubsection{Microwave Spectra}

\begin{figure} 
\centerline{\includegraphics[width=\textwidth]{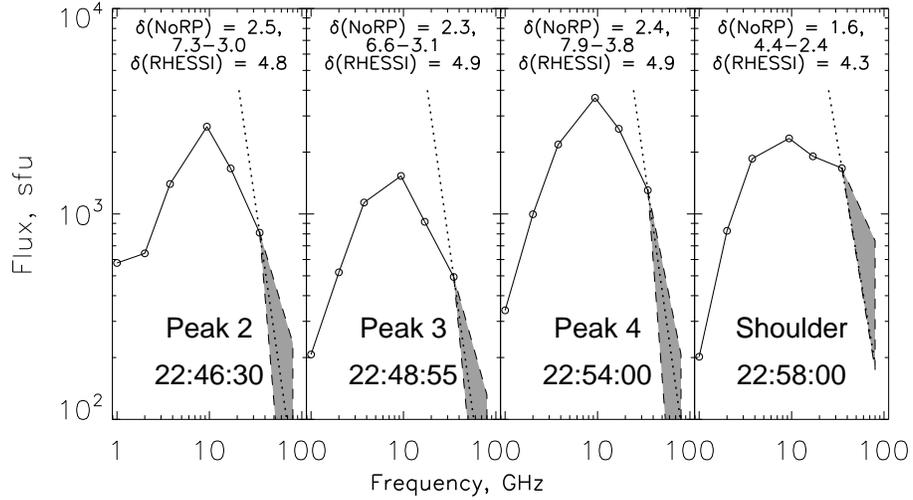}}
\caption{Microwave spectra. Shadings show the uncertainties of the
flux density at 80 GHz. The spectral indices
$\delta_{\mathrm{NoRP}}$ specified in panels a\,--\,d were
calculated from the flux ratios at 17 and 35 GHz (upper row) as
well as from the flux ratios at 35 and 80 GHz at the lower and
upper boundaries of the shaded regions, respectively. The thick
dotted lines show the highest-frequency slope corresponding to the
RHESSI HXR spectrum in the energy range 100\,--\,300 keV produced
by MeV electrons.} \label{F-microwave_spectra}
\end{figure}

Figure~\ref{F-microwave_spectra} shows the microwave spectra at
four different epochs\,---\,peaks 2, 3, 4, and the shoulder, using
the NoRP frequencies at 1.0, 2.0, 3.75, 9.4, 17, 35, and 80 GHz.
All spectra show distinct maxima at about 10 GHz with the
exception of the shoulder which seems to have a complex spectrum,
with two peaks: $\nu_{\mathrm{peak\,1}} \approx 10$~GHz,
$\nu_{\mathrm{peak\,2}} > 20$~GHz. The thick dotted lines in all
panels show the highest-frequency slope $\alpha =
1.22-0.9\,\delta_{\mathrm{RHESSI}}$ corresponding to the RHESSI
spectrum ($\delta_{\mathrm{RHESSI}} = \gamma + 1.5$).

The shapes of the spectra imply non-thermal emission for all four
epochs shown in the figure. Recalling that the microwave flux
densities become almost the same during the decay (see
Figure~\ref{F-rhessi_norp}), it is useful to compare their values
with estimates from soft X-ray observations.

From RHESSI 6\,--\,12 keV images at 22:53\,--\,22:57 we find the
SXR-emitting area to be $A \approx (2-6) \times 10^{18}$~cm$^{2}$
(levels of 0.3\,--\,0.6 of the maximum), and a volume $\sim
A^{3/2} \approx (4-13) \times 10^{27}$~cm$^{3}$. From a thermal
component fit to the RHESSI spectrum below 50 keV at 22:58 we find
an emission measure of $2.8 \times 10^{49} $~cm$^{-3}$ and a
temperature of 20 MK. For comparison, we estimated the temperature
and emission measure variations throughout the event from GOES-12
soft X-ray fluxes assuming coronal abundances \cite{White2005}. At
the same time of 22:58, these estimates provide an emission
measure of $3.5 \times 10^{49} $~cm$^{-3}$, density of $(5-10)
\times 10^{10} $~cm$^{-3}$, and temperature of 15 MK ---
reasonably close to the estimates from RHESSI data. The thermal
radio flux estimated from SXR data is maximum at 22:58 and does
not exceed 28 sfu. Thus, non-thermal emissions appear to dominate
all the microwave/millimeter sources throughout the event.

The slopes of the microwave spectra between the highest
frequencies of 35 and 80 GHz are very uncertain but do not seem to
be inconsistent with electron spectrum power-law indices inferred
from the HXR photon energies in the range 100\,--\,300 keV. The
slopes between 17 and 35 GHz are significantly flatter than the
spectra of the HXR-emitting electrons, with a difference of
2.3\,--\,2.7. This fact is consistent with our preliminary
assumption that the 17 GHz emission is not optically thin,
although the peak frequency of the microwave total flux spectrum
is at about 10 GHz for the most part of the event.

\section{Modeling and Estimates}

\subsection{Microwave/Millimeter Emissions}

To compare the microwave parameters with parameters inferred from
HXR spectra, we use the expressions employed by
\inlinecite{White2003} [based on \inlinecite{Hudson78}] for the 23
July 2002 flare. For a measured photon spectrum of thick-target
non-thermal brems\-strahlung of the form
\begin{eqnarray}
\Phi(E_\gamma) = A_0 \left( \frac {E_\gamma}{E_0}
\right)^{-\gamma} [\mathrm{photons}\ \mathrm{keV}^{-1}\
\mathrm{cm}^{-2}\ \mathrm{s}^{-1}]
 \label{E-ph_powerlaw_spec}
\end{eqnarray}
where $E_\gamma$ is the photon energy, $\gamma$ is the power-law
index of the photon spectrum, and $A_0$ is the normalization
constant at a fiducial photon energy $E_0$~keV (A$_0\,=\,50$ keV
in the OSPEX software used to fit the RHESSI spectra), the energy
distribution of non-thermal electrons takes the form
\begin{eqnarray}
\frac{d^2N(E)}{dE dV} = N_r \frac{\delta-1}{E_r} \left(
{\frac{E}{E_r}} \right)^{-\delta}
 \label{E-Dulk_electron_spectrum}
\end{eqnarray}
(number of electrons of energy $E$ per unit volume and energy)
where $\delta =\gamma+1.5$, $E_r$ is a reference energy (typically
10 keV),  and
\begin{eqnarray}
N_r = 3.04 \times 10^{24}
\frac{A_0b(\gamma)}{(\delta-1)E_{0\,[\mathrm{keV}]}^{0.5}A_X}
\left( {\frac{E_0}{E_r}} \right)^{\delta-1} [\mathrm{electrons}\
\mathrm{cm}^{-3}]
 \label{E-Nr}
\end{eqnarray}
Here $b(\gamma) = \gamma^2(\gamma-1)^2 B(\gamma-0.5,\ 1.5)$ where
$B(x,y)$ is the beta function, and $A_X$ is the area of the HXR
source.

The parameters of the electron energy distribution may be used to
calculate the expected microwave emission due to the
gyrosynchrotron process. There is no simple exact expression for
this calculation, but \inlinecite{DulkMarsh82} provide simple
approximations. The power-law index of the microwave total flux
spectrum in the optically thin limit, $\alpha$, is related to the
power-law index of the electron number spectrum by $\alpha =
1.22-0.9\,\delta$ \cite{DulkMarsh82}. With these expressions it is
possible to compare quantitatively the observed photon HXR spectra
with the microwave spectra.

\subsubsection{Emissions from N1 during Peak 2}

As mentioned, the absence of emission at 34 GHz in the source
above sunspot N1 during Peak 2 might be explained if the 17 GHz
emission is due to a thermal gyroresonance source. As
Figure~\ref{F-hxr_spectra} shows, the HXR spectrum in source N1 is
detectable up to at least 200 keV. Fits to the photon spectrum
show a power-law with $\gamma = 3.3$, but rule out a
high-temperature thermal component.

Estimations of the non-thermal gyrosynchrotron emission from
power-law electrons with $\delta = \gamma +1.5 = 4.8$ using
formulas of \inlinecite{DulkMarsh82} show that the observed 17 GHz
emission above N1 (150 MK) is possible. The ratio of the
brightness temperatures at 34 and 17 GHz is $T_{34}/T_{17}$ $\approx
(34/17)^{(1.22-0.9\delta-2)} \approx 0.029$, and the brightness
temperature expected at 34 GHz in N1 should be about 4.4~MK. The
maximum brightness temperature observed at 34 GHz at peak 2 is
about 90 MK in S2; the dynamic range should be sufficient to
detect a 4.4 MK source (5\%, or $-13$ dB). It is not clear why the
34 GHz emission is absent in N1, if the emission is non-thermal.

A possible solution of this problem is a power-law-like spectrum
with an upper cutoff at a few hundred keV. The number of
moderate-energy electrons is sufficient to produce the observed
strong 17 GHz emission --- even a thermal spectrum would suffice.
However, the deficiency of higher-energy electrons probably
determines the absence of the 34 GHz emission.

\subsubsection{Emissions during Peak 4}

Having parameters of the HXR spectra, one can estimate parameters
of the radio emission. Calculations of radiation of electrons
gyrating in magnetic fields (\textit{e.g.}, \opencite{Ramaty1969};
\opencite{Ramaty1994}; \opencite{PrekaPapadema1992};
\opencite{BBG1998}) appears to be the most rigorous way. Another
way introduced by \inlinecite{DulkMarsh82} uses formal analytic
approximations of the rigorous results. The common problem of both
methods is that there are large uncertainties of several important
parameters; however, the latter method appears to be more
flexible, it is not time-consuming, and allows to get estimates
easily.

\begin{figure} 
\centerline{\includegraphics[width=\textwidth]{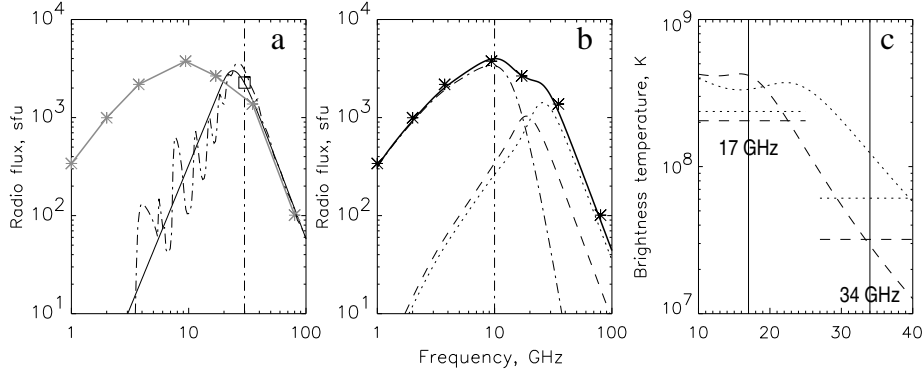}}
\caption{Modeling of the microwave spectrum at Peak 4. (a)~The
observed NoRP spectrum (gray) and the spectra of two identical
footpoint sources modeled using the approach of Dulk and Marsh
(black solid) and the Ramaty code (dash-dotted). (b)~The flux
density spectra of the two footpoint sources (dotted and dashed
lines) and the looptop part (dash-dotted line), and the total
spectrum (thick line). Asterisks in panels (a) and (b) show the
NoRP measurements, and the vertical dash-dotted lines mark the
turnover frequencies. (c)~The spectrum of the brightness
temperatures for the footpoint sources. The vertical lines mark 17
and 34 GHz, and the horizontal lines mark the brightness
temperatures actually observed in these regions.}
 \label{F-spectrum_model}
\end{figure}

First, we try to estimate the magnetic field strength from the
flux density recorded by NoRP at 35 GHz during peak 4 assuming the
emission at this frequency to be optically thin (this seems to be
correct in our case, \textit{e.g.}, according to
Figure~\ref{F-microwave_indices}). Most parameters seem to be
known. The area of the thick target measured from RHESSI
25\,--\,50 keV images was $A_X \approx 1 \times 10^{18}$~cm$^{2}$.
We take $\gamma = 3.4$ [$\delta = 4.9$, $B(\gamma-0.5, 1.5) =
0.16$, and $b(\gamma) = 10.3$], $A_0 \approx 5.2$
photons~cm$^{-2}$~keV$^{-1}$~s$^{-1}$ from the HXR spectrum [see
(\ref{E-ph_powerlaw_spec})]; derive $N_r \approx 3 \times 10^9$
from (\ref{E-Nr}); take the geometrical depth of the emitting
source to be the square root of its area, and the angle between
the line of sight and the magnetic field to be $\approx
60^{\circ}$ according to the position of the active region.
Assuming the total flux to be contributed by two identical
footpoint sources in equal magnetic field strengths, we get $B
\approx 1200$~G from formulas of \inlinecite{DulkMarsh82}. This
seems to be plausible taking account of the estimates of the
photospheric field from the MDI data. With this magnetic field
strength, the frequency maximum is $\nu_{\mathrm{peak}} \approx
30$~GHz, which appears to agree with our preliminary conclusions
made in Section~\ref{S-Radio_Sources}, but is well above the peak
frequency shown by the NoRP total flux spectrum (9.4~GHz). Another
useful quantity is the maximum flux density
$S(\nu_{\mathrm{peak}}) \approx
10^{-19}{k_{\mathrm{B}}\nu_{\mathrm{peak}}^2}/{c^2}\,
\left(1-\exp(-2)\right)T_{\mathrm{eff}}(\nu_{\mathrm{peak}})
\tau(\nu_{\mathrm{peak}})\, \Omega\ \approx 1140\ \mathrm{sfu}$
where $k_{\mathrm{B}}$ is the Boltzmann constant and $\Omega$ is
the solid angle of one source visible from the Earth. The peak
frequency is shown in Figure~\ref{F-spectrum_model}a by the
vertical dash-dotted line, and $2S(\nu_{\mathrm{peak}})$ is shown
by the square. These parameters obviously disagree with the NoRP
spectrum (gray). To understand the situation, we model the
spectrum of the emitting source at a frequency $\nu$ as
\begin{eqnarray}
T(\nu) = T_{\mathrm{eff}}(\nu) [1-e^{-\tau(\nu)}]
 \label{E-t-vs-tau}
\end{eqnarray}
where $T_{\mathrm{eff}}(\nu)$ is the effective temperature of
emitting electrons, $\tau(\nu) = \kappa(\nu)L$ is the optical
thickness, $\kappa(\nu)$ is the absorption coefficient, and $L$ is
the geometrical depth. Both $T_{\mathrm{eff}}(\nu)$ and
$\kappa(\nu)$ are functions of all parameters of the source
determined by formulas of \inlinecite{DulkMarsh82}.

The spectrum modeled in this way and converted to flux density is
shown in Figure~\ref{F-spectrum_model}a by a solid black line. It
significantly differs from the NoRP spectrum. One might assume
that our extension of the approach of \inlinecite{DulkMarsh82} is
not justified, \textit{e.g.}, because we have ignored the fact
that the accuracy of their expressions decreases at high ($>100$)
harmonics of the gyrofrequency and at low ($<10$) ones, with the
latter being more important in our case.

To verify our results, we have overplotted in the same figure the
spectrum calculated using the Ramaty code
\cite{Ramaty1969,Ramaty1994} (dashed line). To co-ordinate the
different geometries of the sources used in both ways, we have
corrected the normalization coefficient for the Ramaty code by a
geometrical factor $k_{\mathrm{geom}} = 4/(3\sqrt{\pi})$, so that
$A_{\mathrm{nor}} = N_r\times (\delta-1)\times
\left(1\,\mathrm{MeV}/E_r \right)^{1-\delta} AL k_{\mathrm{geom}}$
$ = 1.9 \times 10^{29}$ electrons~MeV$^{-1}$. Both model spectra
satisfactorily agree with each other---of course, without
gyroresonance features in the spectrum modeled following
\inlinecite{DulkMarsh82}, which are not really expected in
observations due to inhomogeneity of the magnetic field. Finally,
we note that if asymmetric microwave sources were assumed, then
the magnetic field strength (and the turnover frequency) for one
of them would be still higher.

From comparing the results of the modeling with the observed
spectrum, we conclude that an essential emitting component is
missing, which is minor at high radio frequencies but dominates at
lower frequencies. Even with a frequency-independent brightness
temperature, the area of this component must increase with
wavelength to partially compensate the decrease of $\nu^2$.
Indeed, as known from multi-frequency imaging observations and
modeled theoretically, radio-emitting regions expand with
wavelength. For example, \inlinecite{BBG1998} modeled the radio
emission of a magnetic loop filled with power-law electrons above
a dipole, and the resulting total flux spectrum which they
obtained was broadly similar to our situation. Therefore, besides
the ``kernel'' sources emitting at high microwaves and long
millimeter wavelengths, there must be a larger blob covering them,
with an area and optical thickness increasing with wavelength. We
roughly reproduce the results of \inlinecite{BBG1998} by combining
the radio-emitting regions from two kernel sources localized in
both footpoints of the loop and a blob above them.

We represent the intrinsic brightness temperature of a single source
1 at a frequency $\nu$ according to (\ref{E-t-vs-tau}), and its
issue after the passage through another source 2 as
$\exp(-\tau_2(\nu))$. Thus, the brightness of a kernel source
visible through a blob above it is
\begin{eqnarray}
T(\nu) = T_{\mathrm{eff\,k}}(\nu) \left[
1-e^{-\tau_{\mathrm{k}}(\nu)}\right]e^{-\tau_{\mathrm{b}}(\nu)} +
T_{\mathrm{eff\,b}}(\nu) \left[
1-e^{-\tau_{\mathrm{b}}(\nu)}\right].
\end{eqnarray}
We force the area of the blob $A_{\mathrm{b}}$ to depend on
frequency approximately according to \inlinecite{BBG1998} and
correspondingly change the depth. The magnetic field is also
handled as a growing function of frequency. This approach uses the
fact that while the frequency decreases, a source becomes thicker,
and the contribution from peripheral regions of weaker magnetic
fields grows.

The results of the modeling are shown in
Figure~\ref{F-spectrum_model}b. First of all, we warn against
overinterpreting these results, because the model is coarse,
parameters are not well known, and we therefore did not endeavor
to achieve perfect results. The dotted and dashed lines represent
the ``column'' total flux spectra of the two kernel source visible
through the loop-associated blob, and the dash-dotted line
represents the spectrum of the blob. Panel (c) in the figure shows
the brightness temperatures of the kernels. The vertical lines
mark 17 and 34 GHz, and the horizontal lines mark the brightness
temperatures actually observed in these regions. The relations
between the two sources are roughly reproduced at both
frequencies, although the brightness temperatures at 17 GHz are
higher than the actually observed ones---probably, the sources are
not completely resolved at 17 GHz. The flat parts left from the
turnover frequencies are due to the contributions from the blob.

We used here $A_0 = 5.2$ photons~s$^{-1}$~cm$^{-2}$~keV$^{-1}$
from the observed HXR spectrum, the observed areas of the kernel
sources (see Section~\ref{S-Radio_Sources}), and the magnetic
field strengths  of 1350 and 900 G (stronger in S2, because the
sources were displaced from N1 during peak 4). Their depths were
taken as the square root of the areas. The area of the covering
blob varied from $8\times 10^{18}$~cm$^2$ at 15~GHz up to
$1.5\times 10^{20}$~cm$^2$ at 1~GHz, and its depth was taken to be
$0.2\sqrt{A_{\mathrm{b}}}$. Accordingly, the magnetic field
strength gradually decreased from 540~G to 74~G. Again, we point
out that the values of all quantities are estimates only.

With the coarseness of the model, this exercise nevertheless leads
to the following reasonable conclusions: (\textit{i})~the broad
total flux spectrum could be indeed due to the emission from the
whole loop; (\textit{ii})~the peak frequency shown by NoRP does
not correspond to real turnover frequencies of the main sources
observed at 17 and 34 GHz, being significantly lower; and
(\textit{iii})~the parameters of accelerated electrons found from
HXR spectra appear to correspond to parameters of
microwave-emitting electrons.

Two main results come out from our considerations:
(\textit{i})~flaring in strong magnetic fields, and
(\textit{ii})~inhomogeneity of a microwave source --- in the sense
that different parts of a source dominate its emission at
different frequencies. The major result of our modeling is a
warning for researchers that if the peak frequency observed in the
total flux spectrum is relatively low (\textit{e.g.}, 10 GHz),
this does not guarantee that sources observed at higher
frequencies of 17 GHz or even 34 GHz are optically thin. Our
modeling is rather coarse, but it demonstrates anyway that the
peak frequency observed in the total flux spectrum can be
significantly lower than the peak frequencies in
footpoint-associated sources. This conclusion is consistent with
the assumption of \inlinecite{White2003} and hints at another
possible reason for the long-standing discrepancy between the
power-law indices estimated from HXR and radio data.

\subsubsection{The Shoulder and Decay}

The behavior of the flux density at long millimeter wavelengths
during the shoulder appears to be intriguing (see
Figure~\ref{F-rhessi_norp}): the shoulder is pronounced only at
lower HXR energies ($<300$~keV), which do not significantly affect
microwaves, and its intensity is substantially lower than peak 4,
whereas the opposite situation occurs at 35 and 80~GHz. As the
lower row in Figure~\ref{F-overlay_set} shows, flaring is mainly
concentrated in S2 at that time, which is nearly similar to the
situation during peak 4. The electron spectrum becomes slightly
harder (4.3 against 4.9 at peak 4), but this does not seem to be
sufficient to explain the observations. A possible solution of
this problem might be related to trapping effects. Note also that
the relation between the power-law electron indices inferred from
HXR and microwave spectra observed at peak 4 and the shoulder (see
Figure~\ref{F-microwave_spectra}) hints at progressive hardening
of microwave-emitting electrons with respect to HXR-emitting ones.

\inlinecite{MelroseBrown1976} in their trap-plus-precipitation
model analytically showed that Coulomb collisions in plasma with a
density $n_0$ significantly affect the ``parent'' power-law
electron spectrum with an index $\delta_{\mathrm{inj}}$ injected
into a trap so that the number spectrum of trapped electrons
transforms into a two-part one separated by a transition energy
$E_{\mathrm{T}}$. The $E_{\mathrm{T}}$ moves right with time $t$;
in the non-relativistic limit, $E_{\mathrm{T}} =
(3/2\,\nu_0t)^{2/3}$ with $\nu_0\approx 5\times
10^{-9}n_0$~keV$^{3/2}$~s$^{-1}$. The branches below
$E_{\mathrm{T}}$ and above it depend on the regime of the
injection into the trap. \inlinecite{MelroseBrown1976} considered,
in particular, two limiting injection regimes, i.e., an initial
impulsive injection and a continuous one.
\inlinecite{MetcalfAlexander99} presented in their figures 3 and 4
the spectra calculated for these injection regimes with
$\delta_{\mathrm{inj}} = 4$ for a parent spectrum, which is close
to our case. Schematically, the effects of trapping are as
follows.

After an impulsive injection, the electron number spectrum is
depleted to harden so that the upper envelope of the whole
spectrum goes as $\delta_{\mathrm{inj}}-1.5$, and the branch below
$E_{\mathrm{T}}$ falls as $E^{5/2}$ towards lower energies. During
a continuous injection, the electron number spectrum is augmented
to become a broken double-power-law so that the high-energy branch
keeps a slope of $\delta_{\mathrm{inj}}$ and the low-energy one
takes a slope of $\delta_{\mathrm{inj}}-1.5$. The high-energy
branch augments linearly with time. The spectrum of the HXR
emission produced by electrons precipitating from a trap in the
model of \inlinecite{MelroseBrown1976} is the same as the
thick-target spectrum without trapping due to the steeper spectrum
of the collisional precipitation from a trap. These effects
altogether result in hardening the spectrum of trapped
microwave-emitting electrons with respect to HXR-emitting ones
\cite{MelnikovMagun98}.

The case of a continuous injection resembles the progressive
hardening of microwave emission observed during the shoulder,
especially pronounced in the 80 GHz flux starting from peak 4 by
23:00. The trapping seems to be insignificant at the onset of peak
4, because the polarized emission at 35 GHz closely resembles the
100\,--\,300 keV light curve (see Figure~\ref{F-rhessi_norp}).
Then the 80 GHz flux increases almost linearly with time, as
expected for a continuous injection in a trap.

An additional support in favor of trapping is provided by a
loop-top brightening visible in 34 GHz images obtained during the
shoulder and decay (see, \textit{e.g.}, the upper right image in
Figure~\ref{F-microwave_indices}). The flat spectrum at
9.4\,--\,35~GHz during the decay might be due to combined effects
of trapping and inhomogeneity of the microwave-emitting source.
Both trapped and precipitating electrons can contribute to the
microwave emission \cite{KWS2001} which makes difficult a more
detailed analysis of trapping issues in our event due to
insufficient information.

\section{Discussion and Conclusion}

We have analyzed an event which is interesting in many respects.
It is one of the few events whose high energy emission in hard
X-rays has been mapped above 200 keV; electron spectra inferred
from hard X-rays are consistent with those inferred from microwave
data, but the latter have very large uncertainties. A distinctive
feature of our study is related to strong magnetic fields. The
combination of the MDI calibration correction and the assumption
of radial fields leads magnetic field strengths three times
stronger than uncorrected ones. Such a correction factor appears
to be significant for magnetic fields themselves; it becomes
rather crucial in the interpretation of the gyrosynchrotron
emission. We now illustrate what one would see if the magnetic
fields in footpoint sources in the event were thought to be
2\,--\,3 times weaker.

\begin{enumerate}

\item
 The underestimation of the magnetic field strength results in an
underestimate of the microwave peak frequency; indeed one sees a
low peak frequency of the NoRP total flux spectrum. We have shown
that this low peak frequency is due to the contribution of
emission from the upper blob associated with the whole loop. Thus,
the apparent consistency of the low peak frequency with weak
magnetic field is in this case deceptive.

\item Consequently, radio frequencies which do not correspond to
the optically thin regime can be misinterpreted as belonging to
the optically thin regime. There is no reason to use problematic
80 GHz records in this case, and the microwave spectrum estimated
from the 35 to 17 GHz ratio inevitably becomes flatter than the
optically thin one. The discrepancy with the HXR spectrum then
appears naturally.

\item
 Believing that the 17 GHz emission belongs to the optically thin
regime, one gets a strange behavior of the polarization.

\item
 With the underestimated magnetic field, one gets a significant
deficiency of the flux density, and is constrained to search for a
way to increase it.

\item
 The Razin effect seems to become important at higher frequencies
than in reality. It was most likely negligible in our event.

\end{enumerate}

These considerations might provide a key to reconcile some
puzzling issues established in several other events. As a
by-product of our analysis we conclude that the re-calibrated MDI
magnetograms appear to be more consistent with microwave data than
the previous ones.

Two strong HXR and microwave footpoint sources above sunspots
dominated throughout the event. We find that at all peaks the HXR
sources had nearly the same positions. The flare source in energy
bands from 25 to 400 keV coincided to within 1\arcsec\ at the
onset of peak 4. This suggests that the acceleration process must
be the same for all energy levels --- from 25 to 400 keV.
\inlinecite{GrB08} from their study of the spectral evolution of
HXR bursts also concluded that the observed spectral changes occur
due to ``gradual change in the accelerator'' rather than
contribution from different acceleration mechanisms. Our results
appear to be consistent with this conclusion, although they do not
seem to confirm the model proposed by \inlinecite{GrB08}.

On the other hand, the HXR spectra differ from peaks 2 and 3 to
peak 4. The earlier peaks show a power law with electron index
$\delta\,\approx\,4.9$ from 40 to over 200 keV, with a slightly
flatter (but more uncertain due to low count rates) spectrum above
the break. Peak 4, on the other hand, shows a flatter spectrum
($\delta\,\approx\,4.0$) below photon energies of 120 keV, with
$\delta\,\approx\,4.9$ above this energy. The important result of
spatially resolved spectroscopy in this event is that all three
main sources simultaneously exhibit the flattening in the photon
energy range 40\,--\,120 keV in peak 4: this implies that over the
large volume encompassing these sources, the conditions that
control the spectral index of accelerated electrons changed
simultaneously. This is a severe constraint on acceleration
models.

In terms of magnetic reconnection models (\textit{e.g.},
\opencite{ForbesPriest95}), one might indeed expect the flaring to
be strongest above regions of strong, highly sheared magnetic
fields, and particularly over the umbrae of delta spots. Several
authors (\textit{e.g.}, \opencite{Qiu2002}; \opencite{Asai2004})
have reported a correlation between the energy release rate
computed from footpoint motions across the magnetic fields and
intensities of flare emissions in HXR and microwaves. A similar
correlation probably would exist for the event under discussion.
However, existing flare models do not seem to predict the various
properties of the flare revealed in our paper, \textit{e.g.},
whether a flare would enter a sunspot, how hard the electron
spectrum could be, \textit{etc}.

A class of flares occurring above the sunspot umbrae does not seem
to be sufficiently studied, and their properties have not been
well established. Note that the analysis of the extreme 20 January
2005 event led \inlinecite{Grechnev2008} to a conclusion that its
extremeness was due to the occurrence of the flare above the
sunspot umbrae. One of features observed in that flare was a large
SXR-emitting loop-like structure rooted in the umbrae. Such a loop
between sunspots N1 and S2 was also observed in our 17 June 2003
flare (see Figure~\ref{F-euv_arcade}d). Total magnetic flux was
mainly concentrated between two umbrae, unlike widespread magnetic
fields in a typical flare. This feature seems to be expected:
large total magnetic flux outgoing from a sunspot must be balanced
by incoming flux at the other end of a loop that is favored by the
presence of another sunspot. However, this is one of only a few
expected properties of sunspot-associated flare.

\subsection{Summary}

Our multi-spectral analysis of the 17 June 2003 event has shown
that its main features were probably related to the location of
main flare sources above sunspots. This may determine strong
microwave flare emissions and probably was somehow related to hard
electron spectra observed in the event. Properties of flare
emissions imply a single acceleration mechanism, which was most
likely the same for all energy domains up to 800 keV. Some
features of microwave emissions appear to be indicative of
trapping issues, consistent with existing concepts. We have not
found a significant discrepancy between the spectra of electrons
responsible for microwaves and hard X-rays frequently reported in
previous studies (with the limitation that the microwave index is
very uncertain). Instead, we note that sometimes this discrepancy
could be due to underestimation of the microwave turnover
frequency resulting from inhomogeneity in the microwave/millimeter
source. So we emphasize that the microwave peak frequency measured
from total flux records does not guarantee that higher frequencies
are all optically thin. It rather shows the lower limit of
possible turnover frequencies of gyrosynchrotron spectra of
footpoint-associated sources. This is also related to probable
underestimations of the magnetic field strength. This conclusion
appears to be consistent with the results of
\inlinecite{White2003} which implied an optically thick regime
even at 35~GHz, although their event was significantly different.
These issues highlight the importance of total flux measurements
of radio bursts in the millimeter range. Our results also
emphasize the importance of both experimental and theoretical
analyses of sunspot-associated flares, which might be related to
extreme solar events, but do not appear to be sufficiently
studied.

\acknowledgements

We thank A.M.~Uralov, A.T.~Altyntsev, and H.S.~Hudson for useful
discussions. We thank the instrumental teams of the TRACE mission,
MDI on SOHO, the Nobeyama Solar Facilities, the Big Bear Solar
Observatory, and the GOES satellites. We thank an unknown referee
for useful suggestions.

The research of MRK and SMW for this paper was supported by NSF
grant ATM 02--33907 and NASA grants NAG 5--12860 and
NNG05--GI--91G. The research of EJS was supported by NASA grants
NAG 5--10180 and NNG06--GB--636. The research of VVG and NSM was
supported by the Russian Foundation of Basic Research under grant
07--02--00101.


\end{article}

\end{document}